\documentclass[aps,prd,twocolumn,groupedaddress,nofootinbib,preprintnumbers,eqsecnum,superscriptaddress]{revtex4}

\usepackage{amsmath,amssymb,epsfig,ifthen,capt-of,calc}

\catcode`\à=\active \defà{\`a} \catcode`\À=\active \defÀ{\`A}
\catcode`\á=\active \defá{\'a} \catcode`\Á=\active \defÁ{\'A}
\catcode`\ä=\active \defä{\"a} \catcode`\Ä=\active \defÄ{\"A}
\catcode`\â=\active \defâ{\^a} \catcode`\Â=\active \defÂ{\^A}
\catcode`\å=\active \defå{{\aa}} \catcode`\Å=\active \defÅ{{\AA}}
\catcode`\ç=\active \defç{\c{c}} \catcode`\Ç=\active \defÇ{\c{C}}
\catcode`\è=\active \defè{\`e} \catcode`\È=\active \defÈ{\`E}
\catcode`\é=\active \defé{\'e} \catcode`\É=\active \defÉ{\'E}
\catcode`\ë=\active \defë{\"e} \catcode`\Ë=\active \defË{\"E}
\catcode`\ê=\active \defê{\^e} \catcode`\Ê=\active \defÊ{\^E}
\catcode`\ì=\active \defì{\`{\i}} \catcode`\Ì=\active \defÌ{\`{\I}}
\catcode`\í=\active \defí{\'{\i}} \catcode`\Í=\active \defÍ{\'{\I}}
\catcode`\ï=\active \defï{\"{\i}} \catcode`\Ï=\active \defÏ{\"{\I}}
\catcode`\î=\active \defî{\^{\i}} \catcode`\Î=\active \defÎ{\^{\I}}
\catcode`\ò=\active \defò{\`o} \catcode`\Ò=\active \defÒ{\`O}
\catcode`\ó=\active \defó{\'o} \catcode`\Ó=\active \defÓ{\'O}
\catcode`\ö=\active \defö{\"o} \catcode`\Ö=\active \defÖ{\"O}
\catcode`\ô=\active \defô{\^o} \catcode`\Ô=\active \defÔ{\^O}
\catcode`\ù=\active \defù{\`u} \catcode`\Ù=\active \defÙ{\`U}
\catcode`\ú=\active \defú{\'u} \catcode`\Ú=\active \defÚ{\'U}
\catcode`\ü=\active \defü{\"u} \catcode`\Ü=\active \defÜ{\"U}
\catcode`\û=\active \defû{\^u} \catcode`\Û=\active \defÛ{\^U}
\catcode`\ý=\active \defý{\'y} \catcode`\Ý=\active \defÝ{\'Y}
\catcode`\½=\active \def½{!`}
\catcode`\¾=\active \def¾{?`}
\catcode`\ß=\active \defß{{\ss}}

\newcommand{\tmop}[1]{\ensuremath{\operatorname{#1}}}
\newcommand{\tmem}[1]{{\em #1\/}}
\newcommand{\mathpi}{\pi}
\newcommand{\mathd}{\mathrm{d}}
\newcommand{\mathe}{\mathrm{e}}
\newcommand{\tmscript}[1]{\text{\scriptsize $#1$}}
\newcommand{\bignone}{}
\newcommand{\mathi}{\mathrm{i}}

\begin{document}

\preprint{IFIC/05-68}
\preprint{FTUV/05-1220}
\preprint{HUTP-05/A0053}
\preprint{UG-FT-198-05}
\preprint{CAFPE-68-05}

\title{Geometric approach to condensates in holographic QCD}



\author{Johannes Hirn}
\email{johannes.hirn@ific.uv.es}

\author{Nuria Rius}
\email{nuria@ific.uv.es}
\affiliation{IFIC, Departament de F\'\i sica Te\`orica, CSIC -  Universitat de València, Edifici d'Instituts de Paterna,  Apt. Correus 22085, 46071 València, Spain}

\author{Ver\'onica Sanz}
\email{vsanz@ugr.es}
\affiliation{Jefferson Laboratory of Physics, Harvard University, Cambridge, MA
02138, USA\\
Departamento de F\'\i sica Te\'orica y del Cosmos,  Universidad de Granada, Campus de Fuentenueva, 18071 Granada,  Spain}


\begin{abstract}
An $\tmop{SU} \left( N_f \right) \times \tmop{SU} \left( N_f \right)$
  Yang-Mills theory on an extra-dimensional interval is considered, with
  appropriate symmetry-breaking boundary conditions on the IR brane. UV-brane
  to UV-brane correlators at high energies are compared with the OPE of
  two-point functions of QCD quark currents. Condensates correspond to
  departure from AdS of the (different) metrics felt by vector and axial
  combinations, away from the UV brane. Their effect on hadronic observables
  is studied: the extracted condensates agree with the signs and orders of
  magnitude expected from QCD.
\end{abstract}



\maketitle

\section{Holographic QCD}

Since the pioneering work of {\cite{Son:2003et}}, some attention has been
drawn to what is called {\tmem{holographic QCD}}. This was further
studied using 5D models in
{\cite{Erlich:2005qh,daRold:2005zs,Hirn:2005nr}}. Holographic QCD tries to answer the following
question: {\tmem{If it exists, how does the 5D dual of QCD look
    like?}} The astonishing success of this bottom-up approach warrants a
more detailed study of the qualities the dual should show, independently of a
full-blown stringy set-up {\cite{{Karch:2002sh,Kruczenski:2003uq,Kruczenski:2003be,Babington:2003vm,Evans:2004ia,Nunez:2003cf,Sakai:2004cn,Sakai:2005yt,deTeramond:2005su}}.

Our modus operandi consists of getting a handle on the known behavior of QCD
and learning about the features the dual 5D model should have. At low
energies, one possesses theoretical and experimental information concerning
pion interactions. In the perturbative regime, the OPE of QCD, combined with
the large-$N_c$ expansion and the softness of some amplitudes are powerful
theoretical guidelines for the 5D model's asymptotic behavior. In the
intermediate region, experimental data on spin-1 resonances, like masses and
widths, is available.

Here, we
briefly outline the features of our model.
Our starting point is the fact that gauge fields living in a 5D setup are
supposed to have a 4D global symmetry dual. On the 4D side, QCD possesses a
global $\tmop{SU} \left( N_f \right) \times \tmop{SU} \left( N_f \right)$
symmetry, whose spontaneous breaking gives rise to pions. The main ingredient
of the 5D model is therefore the chiral symmetry of QCD promoted to a 5D gauge
symmetry, as well as its breaking to the diagonal subgroup. A priori, the main
difference between the approach of {\cite{Erlich:2005qh,daRold:2005zs}} and
that of {\cite{Hirn:2005nr}}, is the implementation of the symmetry-breaking.
We discuss later how this influences physical predictions.

In our model we brought into play two ideas related to this breaking. First,
boundary conditions (BCs) in the extra-dimension represent the spontaneous
breaking by a 4D condensate of infinite dimension, before the AdS singularity
has been smoothed out. Second, the gauge sector builds up the pion field: the
Goldstone boson (GB) nature of the latter is thus ensured (protected to all
orders by the 5D gauge symmetry {\cite{Arkani-Hamed:2001ca}}). These two
inputs were enough to provide a massless pion with the right transformation
properties: the model can be rewritten as a 4D lagrangian whose interactions
explicitly obey chiral symmetry (for details, see {\cite{Hirn:2005nr}}). This
ensures that Green's function computed in the model will satisfy the Ward
identities of chiral symmetry.

This picture of chiral symmetry breaking was just the entrée: in the massive
sector, the two infinite towers of vector fields provided by a Kaluza-Klein
decomposition have just the right symmetries to ensure softness of amplitudes
at high energies. The reason behind is the underlying gauge structure which,
from the 4D point of view, manifests itself in the form of clever sum rules
between resonance couplings {\cite{Hirn:2005nr,Sakai:2005yt}}. Moreover, these
two towers provide the partonic logarithm expected from asymptotic freedom, as
a consequence of the asymptotic conformal invariance in the extra-dimension.

Furthermore, the lowest excitations could be identified as the observed $\rho$
and $a_1$ mesons of QCD. Given as inputs the pion decay $f_{\pi}$ constant and
the rho mass $M_{\rho}$, the mass of the $a_1$ was predicted within the
experimental range. Another conspicuous result of the 5D model is that the
experimentally tested rho-meson dominance is an {\tmem{automatic}} property,
whereas it was an {\tmem{assumption}} in 4D models for resonances. It is again
a consequence of the extra-dimensional nature of the model, and manifests
itself in 4D as the decrease of the couplings of the massive resonances to the
pion. Since dominance of the $\rho$ is satisfied within a few percent, and the
soft high-energy behavior is ensured, the 5D models succeed in predicting the
low energy constants of Chiral Perturbation Theory {\cite{Hirn:2005ub}}.

Such 5D models therefore match well to the low-energy predictions of QCD
(which depend on chiral symmetry and lightest-meson dominance), as well as
onto the HE partonic logarithms {\footnote{Note that this is realized with a
fast-growing resonance spectrum: the mass of the $n$-th KK mode goes as $n$
for $n \gg 1$, rather than $\sqrt{n}$ obtained from linear Regge
trajectories.}}. In this work, we ask in addition to reproduce the analytic
form of the first terms in the OPE as given in large-$N_c$ QCD. Various
Ansätze which perform this task for two-point functions can be constructed,
however the 5D approach presents the following appealing features:
\begin{itemize}
  \item There is no need to impose relations between (an infinite number of)
  constants to constrain the HE behavior: the 5D gauge symmetry takes care of
  that task.
  
  \item The condensates (power-corrections to the AdS metric) are input
  parameters in the 5D model, rather than derived quantities. The consequences
  of such deviations from conformality have computable effects on two-point
  and higher Green's functions {\cite{Witten:1998qj,Gubser:1998bc}}.
\end{itemize}
We provide the tools for studying the model in Section \ref{Sect2}, analyze
the results in Section \ref{num} and offer the conclusions in Section
\ref{conclusions}. Appendix \ref{app} proves that our procedure respects
covariance and locality and in Appendix \ref{B} one can find the approximate
dependence of hadronic observables on the condensates.

\section{Incorporating condensates in the 5D model} \label{Sect2}

The simplest model presented in {\cite{Hirn:2005nr}} was useful to illustrate
a basic point: imposing the above-mentioned symmetries (AdS geometry and 5D
gauge $\tmop{SU} ( N_f )_L \times \tmop{SU} ( N_f )_R \rightarrow \tmop{SU} (
N_f )_V$) and adjusting two experimental inputs ($f_{\pi}$ and $M_{\rho}$) was
enough to obtain an agreement for low-energy quantities at the level expected
for a model of large-$N_c$ QCD at leading order.

The next step is to incorporate more features of QCD: the model of
{\cite{Hirn:2005nr}}, defined on truncated AdS space with symmetry-breaking
BCs lacks power corrections in the correlators, known to be present in QCD.
Indeed, the OPE, for large euclidean momentum $Q^2 \equiv - q^2$
{\cite{Shifman:1979bx}} should read, assuming factorization in the large-$N_c$
\begin{eqnarray}
  \Pi_{V, A} \left( - Q^2 \right) & = & - \frac{N_c}{12 \pi^2}  \left\{
  \lambda + \log \left( \frac{Q^2}{\mu^2} \right) +\mathcal{O}( \alpha_s )
  \right\}  \nonumber\\
  & + & \text{$\frac{1}{12 \mathpi}  \frac{\alpha_s  \left\langle G_{\mu \nu}
  G^{\mu \nu} \right\rangle}{Q^4}$} + \begin{pmatrix}  - 1\cr  11 / 7 \end{pmatrix}  \frac{28}{9}  \frac{\mathpi \alpha_s \langle
  \overline{q} q \rangle^2}{Q^6} \nonumber\\
&+&\mathcal{O} \left( \frac{1}{Q^8} \right), 
  \label{2pointgen}
\end{eqnarray}
where $\lambda$ is a subtraction constant. Note that dimension 4 and 6
condensates have quite different chiral properties: the former is
chiral-invariant and therefore does not contribute to $\Pi_V - \Pi_A$, whereas
the latter contributes essentially to $\Pi_V - \Pi_A$ (\ref{2pointgen}). This
is why we will consider both of them. In {\cite{Hirn:2005nr}} the $1 / Q^{2
d}$ pieces of the second line in (\ref{2pointgen}) were absent, and numerical
agreement for $N_c$ was not imposed.

From the model building point of view, there are many ways to introduce the
condensates but, at the end of the day, they are just modifications of the
vector and axial wavefunctions located at a distance from the brane where the
electroweak sector lives {\footnote{In {\cite{Erlich:2005qh,daRold:2005zs}},
couplings of spin-1 fields to a scalar sector were responsible of modifying
the axial wavefunction {\tmem{only}}. No $1 / Q^6$ were generated in the
vector two-point function, in disagreement with (\ref{2pointgen}). Neither
were $1 / Q^4$ terms included, also in contradiction with
(\ref{2pointgen}).}}. Since the model of {\cite{Hirn:2005nr}} already boasts a
pion, there is no need for an extra spin-0 field: we therefore introduce all
condensates as modifications of the metric. We will then be able to originate
{\tmem{both}} axial and vector wavefunction modifications at once in a
parametrically appealing fashion. This will have further consequences in the
low-energy physics, since the modifications will be especially relevant there:
we shall examine them in Section \ref{num}. 

\subsection{Modifications of the metric and OPE}

The extra dimension considered here is an interval. The two ends of the space
are located at $l_0$ (the {\tmem{UV brane}}) and $l_1$ (the {\tmem{IR
brane}}). With a purely AdS metric $w \left( z \right) = l_0 / z$, the
space-time interval then reads in conformal coordinates
\begin{eqnarray}
  \mathd s^2 & = & w \left( z \right)^2  \left( \eta_{\mu \nu} \mathd x^{\mu}
  \mathd x^{\nu} - \mathd z^2 \right) .  \label{2.2}
\end{eqnarray}
This reproduces the logarithm piece in (\ref{2pointgen}), with only
exponentially decaying corrections. Let us see how the power corrections of
(\ref{2pointgen}) can be generated. The high-energy expansion for the two
point functions can be obtained by solving the following differential equation
for a function $\varphi_X \left( z \right)$, $\left( X = V, A \right)$
\begin{eqnarray}
  \left\{ Q^2 - \partial_z^2 - \left( \log w_X \left( z \right) \right)'
  \partial_z \} \varphi_X \left( z \right) \right. & = & 0,  \label{eom}
\end{eqnarray}
where we have used the euclidean 4D momentum $Q^2 = - q^2 > 0$. The boundary
conditions to be imposed on $\varphi_X$ on the IR brane are the same as for
the fields, respectively $\left. \partial_z \varphi_V \left( Q, z \right)
\right|_{z = l_1} = 0$ and $\varphi_A \left( Q, l_1 \right) = 0$, while the
value on the UV brane $\varphi_X \left( Q, l_0 \right)$ is just a
normalization. The two-point function is then obtained through
{\cite{Maldacena:1997re,Gubser:1998bc,Witten:1998qj,Arkani-Hamed:2000ds}}
\begin{eqnarray}
  \Pi_X \left( - Q^2 \right) & = & \left. - \frac{2}{g_5^2}  \frac{1}{Q^2} w_X
  \left( z \right)  \frac{\partial_z \varphi_X \left( Q, z \right)}{\varphi_X
  \left( Q, z \right)} \right|_{z = l_0} .  \label{PiX}
\end{eqnarray}
With a purely AdS metric, $w \left( z \right) = l_0 / z$, one obtains for $Q
\gg 1 / l_1$
\begin{eqnarray}
  &  & \varphi \left( Q, z \right) \simeq \text{$QzK_1 \left( Qz \right)$}, 
\end{eqnarray}
which reproduces the log piece in (\ref{2pointgen}) with the matching
\begin{eqnarray}
  &  & N_c = \frac{12 \mathpi^2 l_0}{g_5^2} .  \label{Nc}
\end{eqnarray}

Assume now that the metrics near the UV take the form
\begin{eqnarray}
  \frac{z}{l_0}  w_{V, A} \left( z \right) & \simeq &  1 + \frac{9
  \mathpi^2}{4 N_c}  \left\langle \mathcal{O}_4 \right\rangle  \left( z - l_0
  \right)^4   \nonumber\\
&+& \frac{5 \mathpi^2}{16 N_c}   \begin{pmatrix}
 -  1\cr
    11 / 7
  \end{pmatrix} \left\langle \mathcal{O}_6 \right\rangle  \left( z -
  l_0 \right)^6 \nonumber\\
&+&\mathcal{O} \left( \left( z - l_0 \right)^8 \right)
  ,  \label{metric2}
\end{eqnarray}
where we have already adopted a convenient name for the coefficients
$\left\langle \mathcal{O}_4 \right\rangle, \left\langle \mathcal{O}_6
\right\rangle$. With the metric (\ref{metric2}), one solves the differential
equation (\ref{eom}) to obtain $\Pi_{V, A}$ via (\ref{PiX})
\begin{eqnarray}
  \Pi_{V, A} \left( - Q^2 \right) & = & - \frac{N_c}{12 \mathpi^2}  \left(
  \log \left( \frac{Q^2}{\mu^2} \right) + \lambda \left( \mu \right) \right) +
  \frac{\left\langle \mathcal{O}_4 \right\rangle}{Q^4} \nonumber\\
&+& \begin{pmatrix}
    - 1\cr
    11 / 7
  \end{pmatrix}  \frac{\left\langle \mathcal{O}_6 \right\rangle}{Q^6}
  +\mathcal{O} \left( \frac{1}{Q^8} \right) .  \label{OPE}
\end{eqnarray}
Deviations from conformality with a given power of $z^{2 d}$ in the metric
(\ref{metric2}) yield a power of $1 / Q^{2 d}$ in the two-point function, with
a computable coefficient. Note that $\mathcal{O}( \alpha_s )$ contributions in
(\ref{2pointgen}) can be accounted as inverse powers of $\log z$ in the metric
{\footnote{We thank Ami Katz for pointing out this fact to us.}}. However, we
restrict our attention to the condensates here.

In the following, we will use the particular functional form
\begin{eqnarray}
  w_X \left( z \right) & = & \frac{l_0}{z} \mathe^{\tmscript{\frac{9
  \mathpi^2}{4 N_c}  \left\langle \mathcal{O}_4 \right\rangle  \left( z - l_0
  \right)^4 + \frac{5 \mathpi^2}{16 N_c}   \begin{pmatrix}
    - 1\cr
    11 / 7
  \end{pmatrix}  \left\langle \mathcal{O}_6 \right\rangle  \left( z -
  l_0 \right)^6}},  \label{exp}
\end{eqnarray}
for which a given power of $z^{2 d}$ in (\ref{metric2}) yields
{\tmem{exactly}} a power of $1 / Q^{2 d}$: only when multiplying (\ref{exp})
by $\exp \left( \frac{12 \mathpi^2}{N_c}  \frac{\Gamma \left( d + 1 / 2
\right)}{\sqrt{\mathpi} d^2 \Gamma \left( d \right)^3} \left\langle
\mathcal{O}_{2 d} \right\rangle  \left( z - l_0 \right)^{2 d} \right)$ is a
term $\left\langle \mathcal{O}_{2 d} \right\rangle / Q^{2 d}$ generated in
(\ref{OPE}).

Representing deviations from conformality as deviations of the metric from AdS
provides a basis to enforce the OPE of QCD term by term. Including the lowest
dimension condensates should be sufficient. Indeed, one would naively expect a
condensate of dimension $2 d$ to be of order the confinment scale to the
appropriate power, i.e. $\Lambda_{\tmop{QCD}}^{2 d}$, and to be proportional
to $N_c$ in the large-$N_c$ limit. Trading $\Lambda_{\tmop{QCD}}$ for the
scale $1 / l_1$ in the 5D model wich induces the mass gap, one thus expects
\begin{eqnarray}
  \langle \mathcal{O}_{2 d} \rangle & \sim & \frac{N_c}{l_1^{2 d}}, 
  \label{estimate}
\end{eqnarray}
or smaller. Plugging this back into (\ref{exp}), one sees that the deviations
of the metric are of order one for the dimension 4 and 6 condensates, but
decrease quite fast with $d$, as $\frac{\Gamma \left( d + 1 / 2 \right)}{d^2
\Gamma \left( d \right)^3}$. Comparing our model to data in Section \ref{num},
we will indeed extract values for condensates that are within the expected
range of (\ref{estimate}) and of QCD phenomenological analyses, confirming the
consistency of the whole approach.

Note that, since it is an order parameter, the dimension-6 condensate
generates a power correction to $\Pi_V - \Pi_A$, implying that there are only
two vanishing Weinberg sum rules. To reproduce this, it is necessary that the
wavefunctions of the vector and axial fields {\tmem{feel}} a different metric.
Otherwise one would have $\left\langle \mathcal{O}_6 \right\rangle_A =
\left\langle \mathcal{O}_6 \right\rangle_V$ in (\ref{OPE}), in contradiction
with the QCD OPE Eq.(\ref{2pointgen}). We explain how this essential point is
achieved in the next Section and Appendix \ref{app}.

\subsection{Definition of the model} \label{lag}

The model is defined by an $\tmop{SU} \left( N_f \right) \times \tmop{SU}
\left( N_f \right)$ YM lagrangian in an extra dimension. We limit ourselves to
terms involving two derivatives. Denoting by $w_{V, A} \left( z \right)$
respectively the warp factors that will be identified later as the metrics
felt by the vector and axial fields, we start from the action
\begin{eqnarray}
  \mathcal{S}_{\text{YM}} & = & - \frac{1}{4 g_5^2}  \int \mathd^4 x
  \int_{l_0}^{l_1} \mathd z \eta^{M N} \eta^{S T} \nonumber\\
  & \times & \left\{ \frac{w_V \left( z \right) + w_A \left( z \right)}{2} 
  \left\langle L_{M S} L_{N T} + R_{M S} R_{N T} \right\rangle \bignone
  \right. \nonumber\\
  & + & \left.  \left( w_V \left( z \right) - w_A \left( z \right) \right) 
  \left\langle L_{M S} \Omega^{\dag} R_{N T} \Omega \right\rangle \right\}, 
  \label{S}
\end{eqnarray}
where the capital letter index runs over the five dimensions. The last term in
(\ref{S}) summarizes the low-energy effects of couplings to other fields, and
will ultimately be responsible for the vector and axial metric being
different. It involves a unitary auxiliary field $\Omega \left( x^{\mu}, z
\right)$, transforming as a bifundamental under the gauge groups. This
spurious field is eliminated by imposing the constraint $D_5 \Omega = 0$.

Note that a term of a similar form, and with similar effects could be written
down in the model of {\cite{Erlich:2005qh,daRold:2005zs}}, using (instead of
the spurion $\Omega$) a bulk scalar $\Phi$. Yet, since $\Phi$ is dimensionful,
this term could only appear at higher orders, presumably suppressed by a large
scale. In our subsequent fit to experimental data, we shall however find that
the order of magnitude of this term is in agreement with our including it at
leading order, i.e. as a deformation of the metric rather than as a
higher-order term. This is directly related to the above estimate of $N_c /
l_1^{2 d}$ as the size of the condensates. Quite the opposite from being a
large scale, $1 / l_1$ is the only scale in the theory: modifications of the
metric are of order one (for the lowest dimension condensates).

Another ingredient in {\cite{Hirn:2005nr}}, responsible for the presence of a
well-defined pion in the spectrum, consists of the choice of BCs. At the IR
boundary $z = l_1$: $L_{\mu} - R_{\mu} = 0$ and $L_{5 \mu} + R_{5 \mu} = 0$.
At the UV boundary, classical configurations serve as sources $\ell_{\mu}
\left( x \right), r_{\mu} \left( x \right)$ for the analogue of the QCD quark
currents, i.e. $L_{\mu} \left( x, z = l_0 \right) = \ell_{\mu} \left( x
\right)$ and $R_{\mu} \left( x, z = l_0 \right) = r_{\mu} \left( x \right)$.
Since the spurion satisfies $D_5 \Omega = 0$, we need only one boundary
condition, and we choose $D_{\mu} \Omega \left( x, z = l_1 \right) = 0$. As
shown by a covariant treatment (Appendix \ref{app}), the action (\ref{S}) is
equivalent to writing a lagrangian in terms of vector and axial field
strengths, but with different metrics, respectively $w_V$ and $w_A$.

\subsection{Sum rules}

In {\cite{Hirn:2005nr}}, four sum rules involving 4D resonances couplings
emerged as a consequence of the underlying 5D gauge structure. Three of them,
together with the relations
\begin{eqnarray}
  L_3 & = & - 3 L_2 \hspace{1em} = \hspace{1em} - 6 L_1, 
\end{eqnarray}
ensured the unsubtracted dispersion relation for the vector form factor, as
well as the once-subtracted dispersion relation for GB forward elastic
scattering amplitudes. These sum rules still hold with the present
modifications of the metric, the same consequences for the behavior of the
amplitudes therefore follow.

The only sum rule that does not hold anymore when $w_A \neq w_V$ is the one
related to the KSFR II ratio: we derive $f_{\pi}^2 / \left( \sum_n g_{V_n}^2
M_{V_n}^2 ) > 3 \right.$ as long as $w_V \left( z \right) - w_A
\left( z \right) < 0$ (as expected in order to reproduce the result of
{\cite{Witten:1983ut}} that $\Pi_V - \Pi_A < 0$). Since this sum rule was the
only one {\tmem{not}} related to any high-energy constraint, this has no
theoretical consequences. From the phenomenological point of view, it is known
that a KSFR II ratio of two is favored, but in a model, this depends on the
predicted values of $f_{\pi}$ and $M_{\rho}$: the agreement is best judged on
the observable $\Gamma_{\rho \rightarrow \pi \pi}$, see Table \ref{tab}.

On the other hand, the case $w_A \neq w_V$ gives rise to a new sum rule
(derived using the completeness relations for the wave-functions of the axial
KK modes)

\begin{eqnarray}
  \sum_{n = 1}^{\infty} f_{A_n}  \left( f_{A_n} + 2 \sqrt{2} \alpha_{A_n}
  \right) & = & 4 \left( L_9 + L_{10} \right),  \label{L910SR}
\end{eqnarray}
in the notation of {\cite{Hirn:2005nr}}. Each term in the left-hand side
vanishes separately when $w_V = w_A$. Due to this sum rule, the axial form
factor $G_A \left( q^2 \right)$ satisfies an unsubtracted dispersion relation
\begin{eqnarray}
  G_A \left( q^2 \right) & = & \sum_{n = 1}^{\infty} f_{A_n}  \left( f_{A_n} +
  2 \sqrt{2} \alpha_{A_n} \right)  \frac{M_{A_n}^2}{M_{A_n}^2 - q^2} . 
  \label{GA}
\end{eqnarray}
The modification $w_V \neq w_A$ implies immediately $L_9 + L_{10} \neq 0$, and
in general $f_{A_n} + 2 \sqrt{2} \alpha_{A_n} \neq 0$ (at least for some $n$).
In addition, since the width $\Gamma_{a_1 \rightarrow \pi \gamma}$ is
proportional to $\left( f_{A_1} + 2 \sqrt{2} \alpha_{A_1} \right)^2$
{\cite{Prades:1993ys,Knecht:2001xc}}, this is accompanied in general by a
non-zero decay of $a_1$ into $\pi \gamma$. For a discussion of the numerical
values of these two quantities, see Appendix \ref{B}.

\subsection{Quark masses} \label{quark}

The electroweak interactions live on the UV brane, which we will therefore
also call the EW brane. These interactions induce masses for the quarks via
the Higgs mechanism of the SM. This in turn breaks the chiral symmetry
explicitly: since we are only modelling QCD, and not electroweak symmetry
breaking, we simply consider hard quark masses. In the present model as in
$\chi$PT {\cite{Gasser:1985gg}}, such quark masses can be introduced via a
spurious field $\chi \left( x \right)$ transforming as a bifundamental under
the chiral $\tmop{SU} \left( N_f \right) \times \tmop{SU} \left( N_f \right)$
symmetry of our 4D world. In our model, this translates as a 4D bifundamental
under a subset of the 5D $\tmop{SU} \left( N_f \right) \times \tmop{SU} \left(
N_f \right)$, namely the one acting on the EW brane. With such an object, one
can write down the following term on the UV boundary
\begin{eqnarray}
  &  & \frac{f_{\pi}^2}{4}  \left\langle \chi \left( x \right) \Omega^{\dag}
  \left( x, l_0 \right) \right\rangle + \text{h.c},  \label{chiOmega}
\end{eqnarray}
where one has to keep in mind that $\chi$ should be counted as equivalent to
an object containing two derivatives, and must be set equal to the quark mass
matrix times a constant $B_0$ (of dimension one).

The key point is then to use the consequences of the constraint $D_5 \Omega =
0$ and the rewritings of Appendix \ref{app}, which lead to $\Omega \left( x,
l_0 \right) = U \left( x \right)$, i.e. the GB matrix. We thus recover the
$\chi$PT term $f_{\pi}^2 / 4 \left\langle \chi U^{\dag} \right\rangle$. From
this it follows, strictly as in $\chi$PT, that $B_0$ gives a measure of the
quark condensate at tree level, according to {\cite{Gasser:1985gg}}
\begin{eqnarray}
  \left\langle \overline{q} q \right\rangle & = & - f_{\pi}^2 B_0 . 
  \label{qbarq}
\end{eqnarray}
As in $\chi$PT again, the term (\ref{chiOmega}) also induces a mass for the
GBs, which become pseudo-Goldstone bosons. Disregarding isospin breaking
effects, this reads {\cite{Gasser:1985gg}}
\begin{eqnarray}
  M_{\pi}^2 & = & B_0  \left( m_u + m_d \right) .  \label{mpi}
\end{eqnarray}
We will use the value $M_{\pi} = 135 \tmop{MeV}$ in the analysis of Section
\ref{num}.

To summarize the situation in the present model, we see that explicit symmetry
breaking by quark masses, which is given by a 4D lagrangian on the EW brane,
works exactly as in 4D $\chi$PT, provided one realizes the equivalence between
$\Omega \left( x, l_0 \right)$ and $U \left( x \right)$. This is also true of
the higher orders.

\section{Numerical results with condensates} \label{num}

Once the model has been rewritten in terms of vector and axial combinations as
in Appendix \ref{app}, one can proceed following the method of
{\cite{Hirn:2005nr}} to compute masses, couplings and decay widths of the
pion, rho and $a_1$. In addition to $f_{\pi}, M_{\rho}, M_{a_1}$, the
following couplings can be easily tested: the coupling of the $\rho$ to the
vector current is tested in $\Gamma_{\rho \rightarrow e e}$, its coupling to
two pions in $\Gamma_{\rho \rightarrow \pi \pi}$, and the couplings of the
$a_1$ to the axial current in $\Gamma_{\tau \rightarrow a_1 \nu_{\tau}}$,
which we compare to the experimental $\text{$\Gamma_{\tau \rightarrow 3 \pi
\nu_{\tau}}$}$. We thus fit data by minimizing the RMS error on the six
quantities $f_{\pi}, M_{\rho}, M_{a_1}, \Gamma_{\rho \rightarrow e e}^{1 / 4},
\Gamma_{\rho \rightarrow \pi \pi}^{1 / 4}, \Gamma_{\tau \rightarrow a_1
\nu_{\tau}}^{1 / 4}$ {\footnote{This is because the decays mentioned involve
the squares of the couplings appearing in the lagrangian: to treat the
couplings we are interested in on the same footing as $f_{\pi}, M_{\rho},
M_{a_1}$ (which appear squared in the lagrangian), we consider the fourth-root
of the decays $\Gamma_{\rho \rightarrow e e}^{1 / 4}, \Gamma_{\rho \rightarrow
\pi \pi}^{1 / 4}, \Gamma_{\tau \rightarrow a_1 \nu_{\tau}}^{1 / 4}$. It is on
these six quantities that we expect the precision to be of the same order.
Note that this procedure is very close to that of {\cite{Erlich:2005qh}},
except that we fit directly observables rather than couplings.}}, with respect
to their central values provided in {\cite{PDG}}. 

We performed a numerical analysis of these quantities in terms of the
condensates and find that the favored region in parameter space is consistent
with the estimate $\left| \left\langle \mathcal{O}_{2 d} \right\rangle \right|
\lesssim N_c l_1^{- 2 d}$. One can then in turn predict the values of the
low-energy constants $L_i$ appearing in the low-energy expansion of the model.

In the favored region, the approximate expression at linear order in the
condensates are accurate within a few percents for $f_{\pi}, M_{\rho},
M_{a_1}$ and the $L_i$ constants. We present these expressions in Appendix
\ref{B}.

\subsection{Strategy and results}

In Ref.{\cite{Hirn:2005nr}}, we followed an approach centered on low-energies:
we imposed $f_{\pi} = 87 \tmop{MeV}$ (in the chiral limit) and $M_{\rho} = 776
\tmop{MeV}$. This led to agreement within the expected range ($1 / N_c$
corrections) for low energy quantities: the mass of the $a_1$, decays of the
$\rho$, and allowed to predict the low-energy constants of $\chi$PT. On the
other hand, this exhausted the free parameters of the model, and yielded a
mismatch in the OPE: the coefficient of the logarithm led to $N_c = 4.3$ in
Eq.(\ref{Nc}) for the chiral limit.

Another way to see this is that a pure AdS model with $N_c = 3$ but without
condensates naturally underpredicts $f_{\pi}$ and the $\chi$PT low-energy
constants (Table \ref{tab}). This mismatch is the evidence that, in order to
interpolate between very high and very low energies with better precision,
another ingredient is needed: the condensates. In the present paper, we start
by imposing $N_c = 3$, as in {\cite{Erlich:2005qh,daRold:2005zs}}. In the pure
AdS case, this leaves us with one free parameter $l_1$, which we adjust to
minimize the RMS error on the six observables $\sqrt{\sum ( \delta O / O )^2 /
n} \bignone$, where n is the number of observables minus parameters ($n = 5$)
\bignone . The best fit is obtained for $l_1 \simeq 3.1 \tmop{GeV}^{- 1}$,
yielding an RMS error of $12.5\%$. This is our input Set A, which constitutes
a benchmark result without condensates: the outputs can be read in Table
\ref{tab}.

The next step is to include the condensates. Dimensional analysis showed that
only the dimension 4 and 6 condensates are relevant. The dimension-4
condensate $\left\langle \mathcal{O}_4 \right\rangle$ accounts for one
parameter since it is chiral-invariant $\left\langle \mathcal{O}_4
\right\rangle \equiv \left\langle \mathcal{O}_4 \right\rangle_V = \left\langle
\mathcal{O}_4 \right\rangle_A$ and the dimension-6 condensates are chosen to
satisfy the relation $\left\langle \mathcal{O}_6 \right\rangle_A = -
\frac{11}{7}  \left\langle \mathcal{O}_6 \right\rangle_V$, derived from
factorization: this adds one more parameter. Figure \ref{ellipse} indicates
approximately the favored region for $\left\langle \mathcal{O}_4
\right\rangle$ and $\left\langle \mathcal{O}_6 \right\rangle$, such that the
RMS error can be below $9 \%$ for an appropriate value of $l_1$. For this
case, $n = 3$.

Inside this favored region, we picked the point denoted by B on the figure.
The outputs for this Set B are given in Table \ref{tab}.




%

  \begin{figure}
  \includegraphics[width=9cm]{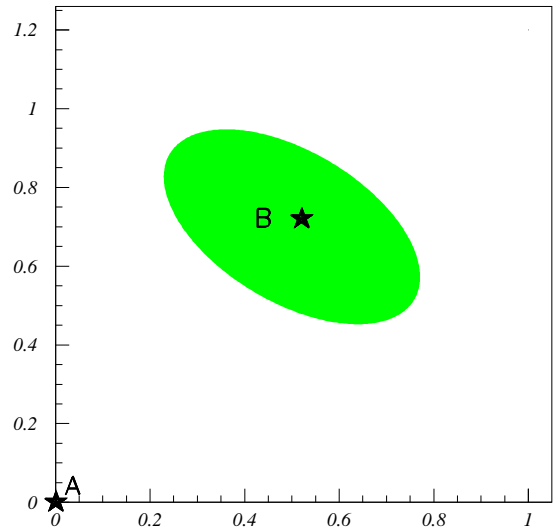}
 \caption{\label{ellipse}Sketch of the favored parameter space: the
   $x$-axis represents $\left\langle \mathcal{O}_4 \right\rangle$ in
   units of $10^{- 3} \tmop{GeV}^4$ and the $y$-axis represents
   $\left\langle \mathcal{O}_6 \right\rangle$ in units of $10^{- 3} \tmop{GeV}^6$. }
  \end{figure}


%
  \begin{table*}
  \caption{\label{tab}Results for the two cases: without and with
  condensates. The signs $\equiv X$ means that the corresponding input
  parameter is fixed to the value $X$.}

  \begin{tabular}{|c||c|c|c|c|}
\hline
 $  $ & $ \text{Set A} $ & $ \text{Set B} $ & $ \text{$\tmop{QCD}$} $ & $ \text{Unit}$\\
  \hline
\hline
  $l_1 $ & $ \equiv 3.1 $ & $ \equiv 3.1 $ & $ \sim 1 / \Lambda_{\text{QCD}} $ & $
  \text{GeV}^{- 1}$\\
  \hline
  $\left\langle \mathcal{O}_4 \right\rangle $ & $ \equiv 0 $ & $ \equiv 0.5 \times
  10^{- 3} $ & $ \frac{1}{12 \mathpi} \alpha_s  \left\langle GG \right\rangle > 0
  $ & $ \text{GeV}^4$\\
  \hline
  $\left\langle \mathcal{O}_6 \right\rangle $ & $ \equiv 0 $ & $ \equiv 0.7 \times
  10^{- 3} $ & $ \frac{28}{9} \mathpi \alpha_s  \left\langle \overline{q} q
  \right\rangle^2 > 0 $ & $ \text{GeV}^6$\\
  \hline
\hline
  $f_{\pi} $ & $ 72.6 $ & $ 84.3 $ & $ 92.4 \pm 0.3 $ & $ \text{MeV}$\\
  \hline
  $M_{\rho} $ & $ 776 $ & $ 796 $ & $ 775.8 \pm 0.5 $ & $ \text{MeV}$\\
  \hline
  $M_{a_1} $ & $ 1.24 $ & $ 1.34 $ & $ 1.230 \pm 40 $ & $ \text{MeV}$\\
  \hline
  $\Gamma_{\rho \rightarrow e e} $ & $ 5.63 $ & $ 5.80 $ & $ 7.02 \pm 0.11 $ & $ \text{KeV}$\\
  \hline
  $\Gamma_{\rho \rightarrow \pi \pi} $ & $ 160 $ & $ 124 $ & $ 150.3 \pm 1.6 $ & $ \text{MeV}$\\
  \hline
  $\Gamma_{\tau \rightarrow a_1 \nu_{\tau}} $ & $ 0.563 $ & $ 0.580 $ & $  0.414 $ & $
  \text{meV}  $\\
\hline
  \hline
  $10^3 L_1 $ & $ 0.36 $ & $ 0.45 $ & $ 0.4 \pm 0.3 $ & $ $\\
  \hline
  $10^3 L_2 $ & $ 0.73 \left( = 2 L_1 \right) $ & $  0.89 \left( = 2 L_1 \right) $ & $ 1.35 \pm 0.3 $ & $ $\\
  \hline
  $10^3 L_3 $ & $ - 2.2 \left( = -6 L_1 \right)  $ & $ - 2.7 \left( = -6 L_1 \right) $ & $ - 3.5 \pm 1.1 $ & $ $\\
  \hline
  $10^3 L_9 $ & $ 4.7 $ & $ 5.5 $ & $ 6.9 \pm 0.7 $ & $ $\\
  \hline
  $10^3 L_{10} $ & $  - 4.7 \left( = - L_9 \right)$ & $ - 5.4 $ & $ - 5.5 \pm 0.7 $ & $ $\\
  \hline
\end{tabular}
 \end{table*}


In Table \ref{tab}, we indicate the result for low-energy observables, without
condensate (Set A) and with condensates (Set B). We remind the reader of the
correspondance between our parameters and QCD condensates. We present the
values of the six low-energy observables taken into account in the fit, as
compared to the values extracted from {\cite{PDG}}. We also give the
corresponding predictions for the low-energy constants of $\chi$PT, as
compared to the $\mathcal{O} \left( p^4 \right)$ fit of
{\cite{Bijnens:1994qh}} {\footnote{At the level we are working, it does not
make sense to consider $\mathcal{O} \left( p^6 \right)$ fits.}}, renormalized
at the scale $\mu = 770 \tmop{MeV}$. Note that the presence of positive
$\left\langle \mathcal{O}_4 \right\rangle$ and $\left\langle \mathcal{O}_6
\right\rangle$ essentially improves the prediction for $f_{\pi}$ and the $L_i$
constants, as is readily seen from the approximate expressions of Appendix
\ref{B}. In summary, condensates improve the matching between very low
energies ($f_{\pi}$ and $L_i$'s of $\chi$PT) on one hand, and very high
energies ($N_c = 3$) on the other hand.

\subsection{Interpretation}

In such a model of large-$N_c$ QCD treated at leading order, we expect
corrections in $1 / N_c$ to parameters in the lagrangian. To facilitate
comparison with {\cite{Erlich:2005qh}}, we have considered square root of such
parameters here. Still, experience shows that large-$N_c$ models usually fare
better than this rough estimate, except for some observables in which
disagreement is expected (an example here is $\Gamma_{a_1 \rightarrow \pi
\gamma}$, see Appendix \ref{B}, where the related case of $L_9 + L_{10}$ is
also discussed).

Our benchmark Set A, with vanishing condensates, already has an RMS error of
12.5\%. Still, our analysis indicates a preference for non-vanishing
condensates: Set B has an RMS error of 8.5 \% . Fixing the best fit involves
just low-energy quantities such as decays and masses, but from this set of
observables one can infer information about the high energy. The output is
that the preferred signs for the condensates agree with those predicted by
QCD: $\left\langle \mathcal{O}_4 \right\rangle > 0$ and $\left\langle
\mathcal{O}_6 \right\rangle > 0$. In fact, the absolute value of $\left\langle
\mathcal{O}_4 \right\rangle$ is consistent with the values extracted from QCD
phenomenology (see for instance
{\cite{Yndurain:1999pb,Ioffe:2002be,Davier:2005xq}}, and references therein).

As for the dimension-6 condensate, we can estimate its size using
factorization. However, since we have not explicitly introduced the quark
masses and the strong coupling constant into the model, $\alpha_s 
\left\langle \overline{q} q \right\rangle^2$ can only be obtained from $m_q 
\left\langle \overline{q} q \right\rangle$ using input from QCD. One starts
from the Gell-Mann-Oakes-Renner relation (Eqs.(\ref{qbarq}-\ref{mpi})), giving
a value of $m_q  \left\langle \overline{q} q \right\rangle$. Using knowledge
of the quark masses and including corrections, {\cite{Jamin:2002ev}} finds
$\left\langle \overline{q} q \right\rangle \left( 2 \tmop{GeV} \right) =
\left( 0.267 \pm 0.016 \text{ } \tmop{GeV} \right)^3$. This yields $28 / 9
\mathpi \alpha_s  \left\langle \overline{q} q \right\rangle^2 \sim 10^{- 3}
\tmop{GeV}^6$, with quite large errors that encompass our favored values of
$\left\langle \mathcal{O}_6 \right\rangle$ (Table \ref{tab} and Figure
\ref{ellipse}). Note that phenomenologial QCD studies 
(see for instance
    {\cite{Davier:1998dz,Ioffe:2000ns,Cirigliano:2003kc,Rojo:2004iq,Zyablyuk:2004iu,Friot:2004ba,Narison:2004vz,Bordes:2005wv}} and references therein)
find in general a larger value for the dimension-6 condensate, more difficult
to reconcile with factorization.

Finally, note that the favored region we have found corresponds to an approximate cancellation between
the effects of $\left\langle \mathcal{O}_4 \right\rangle$ and
$\left\langle \mathcal{O}_6 \right\rangle$ on $w_V$, while they add up in the axial
channel. This supports the succes of the model of
{\cite{Erlich:2005qh,daRold:2005zs}}, where condensates were
introduced in the axial channel only, via a bulk scalar field.

\section{Conclusions} \label{conclusions}

We have started from the simplest model of holographic QCD and described a way
to implement term by term the OPE of QCD. The 5D model we start with is the
simplest in the sense that it  is defined in terms of Yang-Mills fields only,
with appropriate BCs that implement the spontaneous symmetry-breaking. This
model has only two free parameters: a scale given by the inverse of the
extension of the extra dimension, and a parameter to be identified with the
number of colors of QCD.

In AdS, distance in the extra dimension measures the 4D momentum: each
condensate, giving an giving a power-correction in $1 / Q^{2 d}$ in the deep
euclidean, corresponds in the 5D picture to a power-correction $z^{2 d}$ to
the AdS metric, where $z$ measures the separation from the UV brane. We use a
formalism involving a spurion, equivalent to considering two different metrics
for the vector and axial fields. This describes order parameters in the QCD
language.

The expansion in modifications of the metric is under control. First, we find
that the naive estimate for its size is indeed verified by data and secondly,
the effect of these modifications become irrelevant as the dimension of the
condensate increases. This justifies a posteriori our procedure of including
deformations of the metric rather than higher-order operators.

By introducing the modifications of the metric, we show that the infinite
number of Weinberg sum rules are reduced to the desired number, i.e. two. Also
note that the high-energy properties of the amplitudes studied in
{\cite{Hirn:2005nr}} are not lost, and in fact, a similar property for the
axial form factor can be demonstrated.

In terms of input parameters, this model with condensates adds two: the
dimension 4 and 6 condensates $\langle \mathcal{O}_4 \rangle_V = \left\langle
\mathcal{O}_4 \right\rangle_A$ and $\langle \mathcal{O}_6 \rangle_V = - 7 / 11
\left\langle \mathcal{O}_6 \right\rangle_A$. We performed a complete numerical
analysis of the effect of those condensates on hadronic observables: $f_{\pi},
M_{\rho}, M_{a_1}, \Gamma_{\rho \rightarrow e e}^{1 / 4}, \Gamma_{\rho
\rightarrow \pi \pi}^{1 / 4}$ and $\Gamma_{\tau \rightarrow a_1 \nu_{\tau}}^{1
/ 4}$. We find that the presence of the condensates improves the agreement
from 12.5 \%  to 8.5 \% , even though the number of parameters has increased
from1 to 3. Moreover, the favored region in the parameter space corresponds to
non-zero values for both dimension 4 and 6 condensates. The specific value for
$\langle \mathcal{O}_4 \rangle$, corresponding to $1 / ( 12 \mathpi ) \alpha_s
\left\langle G_{\mu \nu} G^{\mu \nu} \right\rangle$, is of $\left( 0.2 \div
0.8 \right) \times 10^{- 3} \tmop{GeV}^4$. For $\langle \mathcal{O}_6
\rangle$, corresponding to $28 / 9 \mathpi \alpha_s  \left\langle \overline{q}
q \right\rangle^2$, the preferred range is $\left( 0.4 \div 1.0 \right) \times
10^{- 3} \tmop{GeV}^6$, in agreement with the signs and orders of magnitude
expected from QCD and factorization for the dimension-6
condensate.

We also extracted the corresponding low-energy constants: condensates improve
the agreement. In particular, both $L_9 + L_{10}$ and $\Gamma_{a_1 \rightarrow
\pi \gamma}$ become non-zero, in connection with the presence of a dimension-6
condensate. However, if factorization is not largely violated, both quantities
remain smaller than the experimental value.

Apart from the already mentioned differences between our model and the one by
{\cite{Erlich:2005qh,daRold:2005zs}}, we mention a more theoretical one: in the case of
{\cite{Erlich:2005qh,daRold:2005zs}}, the massless pion is an admixture
containing a bulk scalar, not protected by the gauge symmetry. In their
construction, the limit of vanishing {\tmem{condensates}} implies the
restoration of chiral symmetry in the spectrum: equal masses for the vector
and axial resonances, and the absence of a massless pseudo-scalar. In our
case, vanishing of {\tmem{local}} order parameters does {\tmem{not}} imply
vanishing of {\tmem{non-local}} order parameters, and therefore chiral
symmetry is still broken (the axial masses are distinct from the vector ones
and the Goldstone bosons remain in the spectrum, with $f_{\pi} \neq 0$, see
Appendix \ref{B}). The only way to reinstate chiral symmetry in the spectrum
is then to send $f_{\pi}$ to zero. In practice, this requires $l_1 \rightarrow
\infty$, i.e. the vanishing of the mass gap: a continuum of states is
recovered. The converse of this statement, namely that confinment implies
chiral symmetry breaking, is therefore built into our model.

\begin{acknowledgments}
We thank Leandro da Rold, Ami Katz, Toni Pich, Alex Pomarol, Jorge
Portol\'es, Jan Stern and Pere Talavera for discussions.
This work was partially
supported by the Spanish MCyT grants BFM2002-00345 and
FPA2004-00996, by the EC RTN networks MRTN-CT-2004-503369 and
HPRN-CT-2002-00311 and by the Generalitat Valenciana grants GV04B-594,
GV05/015, GV05/264  and GRUPOS03/013.
\end{acknowledgments}

\appendix
\section{Symmetry-breaking, covariance and locality} \label{app}

In the action (\ref{S}), we introduced a term $\left\langle L_{M N}
\Omega^{\dag} R_{S T} \Omega \right\rangle$. It is clear that such a crossed
term is going to affect the vector and axial channels differently, and thus
introduce symmetry-breaking effects. The role of the auxiliary 5D field
$\Omega$ is the following: one cannot simply write down $\left\langle L_{M N}
R_{S T} \right\rangle$, because this term does not respect the 5D $\tmop{SU}
\left( N_f \right) \times \tmop{SU} \left( N_f \right)$ group. To make it
invariant, it appears that we need Wilson lines going from the point $z$ to
$l_1$, but this would violate locality in the fifth dimension. This would be
forgetting one thing: to make the whole procedure covariant, we should in
addition reinstate the coset elements on the IR brane, which belong to the
group $\left[ \tmop{SU} \left( N_f \right) \times \tmop{SU} \left( N_f \right)
\right] / \tmop{SU} \left( N_f \right)$ acting at $z = l_1$. The question is
then the following: is there a way to write the appropriate combination of
these elements as a local object?

The answer to the above turns out to be yes, as we show here. For simplicity,
we start from the solution to the problem, and derive the result: we introduce
an auxiliary field $\Omega \left( x, z \right)$ in the bulk, without kinetic
term. Provided the right constraints are applied, this object can be
decomposed as the above-mentioned product of Wilson lines and IR-brane coset
elements. Therefore, a spurion $\Omega$ with a constraint solves the
difficulty mentioned above, by allowing a writing that is both covariant and
local. It is only by solving the constraint that the Wilson lines (and the
coset element) appear. This we show below, as well as the rewriting in the
form of two different metrics for (covariant) vector and axial combinations.
As an intermediate step, we first reinstate the coset elements on the IR
brane.

To do this, we go back to the question of symmetry breaking by BCs. Rather
than directly identifying connections on the IR brane, we use a spurion on the
IR brane to render the formalism covariant. This amounts to performing the
identification up to a gauge, as done in {\cite{Hirn:2005fr}}. In the present
case, one would introduce a field $\omega \left( x \right)$ living on the IR
brane, on which the constraint $D_{\mu} \omega = 0$ is applied. $\omega$
should in fact be an element of the coset of the transformations on the IR
brane, i.e. it belongs to $\left[ \tmop{SU} \left( N_f \right) \times
\tmop{SU} \left( N_f \right) \right] / \tmop{SU} \left( N_f \right)$. In other
words, it is unitary, and transform as a bi-fundamental under the $\tmop{SU}
\left( N_f \right) \times \tmop{SU} \left( N_f \right)$ gauge group. We impose
the two constraints
\begin{eqnarray}
 0 &= & D_{\mu} \omega \left( x \right) \nonumber\\
&=&\partial_{\mu}
  \omega \left( x \right) - \mathi R_{\mu} \left( x, l_1 \right) \omega \left(
  x \right) + \mathi \omega \left( x \right) L_{\mu} \left( x, l_1
  \right), \\
0 &=&  L_{5 \mu} \left( x, l_1 \right) + \omega^{\dag} \left( x \right) R_{5 \mu}
  \left( x, l_1 \right) \omega \left( x \right),
\end{eqnarray}
which are both invariant under the 5D $\tmop{SU} \left( N_f \right) \times
\tmop{SU} \left( N_f \right)$ gauge group. They provide a covariant
replacement for the BCs on the IR brane.

This defines the model as well as BCs: one can then choose to gauge away the
coset element $\omega$, the rest of the algebra is then strictly identical to
that of {\cite{Hirn:2005nr}}. Alternatively, one may want to preserve
covariance all along, in which case the definitions should be slightly
modified: for instance, the rotations $\xi_{L, R} \left( x, z \right)$
appearing in the field redefinitions used to obtain appropriate vector and
axial fields $\hat{V}_M, \hat{A}_M = \frac{\mathi}{2} \left\{ \xi_L^{\dag}  \left(
\partial_M - \mathi L_M \right) \xi_L \pm \xi_L^{\dag}  \left( \partial_M -
\mathi L_M \right) \xi_L \right\}$ should involve Wilson lines, but also the
field $\omega$. The constraint to respect is that

\begin{eqnarray}
\xi_R \left( x, z \right)
\xi_L^{\dag} \left( x, z \right) &=& \text{P} \left\{ \exp \left( \mathi
\int_{l_1}^z \mathd x^5 R_5 \left( x, x^5 \right) \bignone \right) \right\}
\omega \left( x \right)  \nonumber\\
&\times& \text{P} \left\{ \exp \left( \mathi \int_z^{l_1}
\mathd x^5 L_5 \left( x, x^5 \right) \bignone \right) \right\} .
\end{eqnarray}
With this
definition, the matrix $U \left( x \right)$ collecting the pions is still
written as $U = \xi_R \left( x, l_0 \right) \xi_L^{\dag} \left( x, l_0
\right)$, and all equations follow identically as in {\cite{Hirn:2005nr}}, but
now preserving the full covariance at each step.

This shows how a 4D spurion on the IR brane reintroduces the coset element
necessary for a covariant description of the symmetry-breaking conditions at
the boundary. Note that the original lagrangian was local, but our field
redefinitions are non-local in the fifth dimension, since they involve Wilson
lines: this will always be the case in such a model where the GB mode
{\tmem{is}} the Wilson line, and therefore non-local. We now move to the
question of the bulk auxiliary field $\Omega \left( x, z \right)$: we start
from a local and covariant lagrangian generalizing the previous idea. We have
already mentioned the constraint to be imposed
\begin{eqnarray}
0&=&  D_5 \Omega \left( x, z \right) \nonumber\\
& = & \partial_5 \Omega \left( x, z \right) -
  \mathi R_5 \left( x, z \right) \Omega \left( x, z \right)
  \nonumber\\
&+& \mathi \Omega
  \left( x, z \right) L_5 \left( x, l_1 \right),
  \label{cnstrnt}
\end{eqnarray}
as well as the BC for $\Omega$
\begin{eqnarray}
  D_{\mu} \Omega \left( x, z = l_1 \right) & = & 0,  \label{BCW}
\end{eqnarray}
when we wrote down the action (\ref{S}). Let us now describe how, starting
from a lagrangian that respects locality in the fifth dimension, as well as
covariance under the 5D $\tmop{SU} \left( N_f \right) \times \tmop{SU} \left(
N_f \right)$ gauge group, the constraints above induce two different metrics
for vector and axial field strengths.

The solution to the constraint (\ref{cnstrnt}) can be written as 
\begin{eqnarray}
\Omega
\left( x, z \right) &=& \text{P} \left\{ \exp \left( \mathi \int_{l_1}^z \mathd
x^5 R_5 \left( x, x^5 \right) \bignone \right) \right\} \Omega \left( x, l_1
\right) \nonumber\\
&\times& \text{P} \left\{ \exp \left( \mathi \int_z^{l_1} \mathd x^5 L_5
\left( x, x^5 \right) \bignone \right) \right\} .
\end{eqnarray}
Using the BC (\ref{BCW})
then brings us back to the previous case, with the replacement $\omega \left(
x \right) \longmapsto \Omega \left( x, l_1 \right)$, and all the equations
follow. In particular, we see that
\begin{eqnarray}
  &  & \left( w_V \left( z \right) + w_A \left( z \right) \right) 
  \left\langle L_{M S} L_{N T} + R_{M S} R_{N T} \right\rangle
  \bignone \nonumber\\
&+& 2
  \left( w_V - w_A \right)  \left\langle L_{M N} \Omega^{\dag} R_{S T} \Omega
  \right\rangle \nonumber\\
  & = & 4 w_V  \left\langle \hat{V}_{M N}  \hat{V}_{S T} \right\rangle + 4 w_A 
  \left\langle \hat{A}_{M N}  \hat{A}_{S T} \right\rangle, 
\end{eqnarray}
confirming the interpretion that this term splits the two metrics, at the
level of the (covariant) vector and axial combinations of fields. From here
on, the extraction of the lagrangian in terms of the pions $U$ and the KK
modes $V_n, A_n$ proceeds as in {\cite{Hirn:2005nr}}. The pion field is
defined again by $U \left( x \right) = \xi_R \left( x, l_0 \right)
\xi_L^{\dag} \left( x, l_0 \right)$, which is also equal to $\Omega \left( x,
l_0 \right)$, hence the straightforward identification for the quark mass
terms of Section \ref{quark}.

\begin{widetext}
\section{Approximate expressions} \label{B}

We give approximate expressions for some quantities, at linear order in the
condensates. Using reduced values of the condensates $o_4 \equiv \left\langle
\mathcal{O}_4 \right\rangle / ( N_c l_1^{- 4} )$ and $o_6 \equiv \left\langle
\mathcal{O}_6 \right\rangle / \left( N_c l_1^{- 6} \right)$, we have
{\footnote{The numerical coefficients stem from integrals of Bessel
functions.}}
\begin{eqnarray}
  f_{\pi} & = & \frac{\sqrt{N_c}}{\sqrt{6} \mathpi l_1}  \left( 1 + \frac{3
  \mathpi^2}{8} o_4 + \frac{55 \mathpi^2}{896} o_6 \right) +\mathcal{O} \left(
  o^2 \right), \\
  M_{\rho} & \simeq & \frac{2.40}{l_1}  \left( 1 - 3.35 o_4 + 0.38 o_6 \right)
  +\mathcal{O} \left( o^2 \right), \\
  M_{a_1} & \simeq & \frac{3.83}{l_1}  \left( 1 + 1.01 o_4 + 0.29 o_6 \right)
  +\mathcal{O} \left( o^2 \right), \\
  L_3 \hspace{1em} = \hspace{1em} - 3 L_2 \hspace{1em} = \hspace{1em} - 6 L_1
  & = & - \frac{11 N_c}{1536 \mathpi^2}  \left( 1 + \frac{3 \mathpi^2}{2} o_4
  + \frac{145 \mathpi^2}{8624} o_6 \right) +\mathcal{O} \left( o^2 \right), 
  \label{L3}\\
  L_9 & = & \frac{N_c}{64 \mathpi^2}  \left( 1 + \frac{25 \mathpi^2}{24} o_4 +
  \frac{5 \mathpi^2}{448} o_6 \right) +\mathcal{O} \left( o^2 \right), 
  \label{L9}\\
  L_{10} & = & 
    - \frac{N_c}{64 \mathpi^2}  \left( 1 + \frac{25 \mathpi^2}{24} o_4 -
    \frac{3 \mathpi^2}{448} o_6 \right) +\mathcal{O} \left( o^2 \right) . 
    \label{L10}
\end{eqnarray}
It turns out that the expressions are accurate to a few percent for the range
of interest (the one of Figure \ref{ellipse}).
\end{widetext}
Adding (\ref{L9}) and (\ref{L10}) shows that, if $\left\langle \mathcal{O}_6
\right\rangle$ is of the order of magnitude expected from factorization on
obtains  $L_9 + L_{10} \sim 2 \times 10^{- 4}$, i.e. one order of magnitude
below the experimental value. Similarly, the model underpredicts the decay
$\Gamma_{a_1 \rightarrow \pi \gamma}$ if factorization is respected. This can
be understood as follows: a model that reproduces the width $\Gamma_{\tau
\rightarrow a_1 \nu_{\tau}}$ has $f_{A_1} \sim 0.13$. Assuming that the sum
rule (\ref{L910SR}) is saturated by the $a_1$ one can estimates $\Gamma_{a_1
\rightarrow \pi \gamma}$ from the knowledge of $L_9 + L_{10}$ from
factorization. A suppression by an order of magnitude for $L_9 + L_{10}$
yields a suppression by two orders of magnitude for $\Gamma_{a_1 \rightarrow
\pi \gamma}$. Indeed, we find that $\Gamma_{a_1 \rightarrow \pi \gamma} \sim
\tmop{few} \tmop{keV}$ for the favored region of Figure \ref{ellipse}, whereas
experimentally one has $\Gamma_{a_1 \rightarrow \pi \gamma} = 640 \pm 246
\tmop{keV}$ {\cite{PDG}}.



\bibliography{bib-biblio}

\end{document}